\documentclass[trackchanges]{aastex7}
\usepackage{amsmath}
\usepackage{algorithm}
\usepackage{algpseudocode}

\newcommand{\CIERA}{Center for Interdisciplinary Exploration and Research in Astrophysics,
Northwestern University, 1800 Sherman Avenue, Evanston, Illinois 60201, USA}
\newcommand{\SKAI}{The NSF-Simons AI Institute for the Sky (NSF-Simons SkAI), 875 N. Michigan Ave., Suite 4010, Chicago, IL 60611}

\begin{document}

\title{Emulating compact binary population synthesis simulations with uncertainty quantification and model comparison using Bayesian normalizing flows.}

\author[0000-0002-7322-4748]{Anarya Ray}
\email[show]{anarya.ray@northwestern.edu}
\affiliation{\CIERA}
\affiliation{\SKAI}

\begin{abstract}
Population synthesis simulations of compact binary coalescences~(CBCs) play a crucial role in extracting astrophysical insights from an ensemble of gravitational wave~(GW) observations. However, realistic simulations can be costly to implement for a dense grid of initial conditions. Normalizing flows can emulate population synthesis runs to enable simulation-based inference from observed catalogs and data augmentation for feature prediction in rarely synthesizable sub-populations. However, flow predictions can be wrought with uncertainties, especially for sparse training sets. In this work, we develop a method for quantifying and marginalizing uncertainties in the emulators by implementing the Bayesian Normalizing flow, a conditional density estimator constructed from Bayesian neural networks. Using the exact likelihood function naturally associated with density estimators, we sample the posterior distribution of flow parameters with suitably chosen priors to quantify and marginalize over flow uncertainties. We demonstrate the accuracy, calibration, inference, and data-augmentation impacts of the estimated uncertainties for simulations of binary black hole populations formed through common envelope evolution. We outline the applications of the proposed methodology in the context of simulation-based inference from growing GW catalogs and feature prediction, with state-of-the-art binary evolution simulators, now marginalized over model and data uncertainties.

\end{abstract}



\section{Introduction}

The origin of merging compact binaries~(CBCs) observed by ground-based gravitational wave~(GW) detectors~\citep[such as the LVK, ][]{LIGOScientific:2014pky,VIRGO:2014yos,KAGRA:2020agh} remains an open question~\citep{Mandel:2018hfr, Mapelli:2018uds, Mapelli:2020vfa}. Several formation scenarios have been proposed in existing literature, most of which are characterized by partially unconstrained physics. Broadly, CBCs capable of merging within a Hubble time are expected to emerge from isolated evolution of massive stellar binaries that undergo orbital hardening, via either common-envelope, stable mass transfer, or chemical mixing~\citep[e.g., ][]{Postnov:2006hka,1976IAUS...73...75P,vandenHeuvel:2017pwp, Marchant:2016wow,Mandel:2015qlu}, dynamical assembly assisted by either a tertiary companion, multiple exchanges in dense clusters, or gas-assisted migration~\cite[e.g.,][]{Wen:2002km,Antonini:2012ad,Benacquista:2011kv,Bartos:2016dgn}, and hierarchical mergers in star clusters and AGN disks~\citep{Gerosa:2021mno}. The uncertain initial conditions and parameters that underlie these evolutionary pathways are crucial to elucidating unconstrained astrophysical phenomena and are often imprinted on the measurable population properties of CBC observations.

Population synthesis~\citep[PopSynth, see ][ for a review]{breivik2025populationsynthesisgravitationalwave} simulations can model these astrophysical imprints on the various features and correlations of the population-level distributions of measurable CBC parameters. PopSynth predictions can guide the construction of targeted phenomenological models~\citep[e.g.,][]{Talbot:2018cva, vanSon:2021zpk} of the distribution function whose parameters can be constrained given GW data from growing catalogs to extract astrophysical insights~\citep{KAGRA:2021duu}. More flexible population models~\citep[e.g.,][]{Edelman:2022ydv, Callister:2023tgi, Ray:2023upk, Ray:2024hos, Heinzel:2024jlc} are also employed to search for additional physics in the data beyond the assumptions of astrophysically motivated density functions, with the reconstructed population features being harder to interpret in terms of the physical processes underlying an evolutionary channel. On the other hand, direct comparison of PopSynth outputs to a collection of measured CBC parameters by means of density estimation can provide observational constraints on astrophysical initial conditions of the simulated evolutionary pathways~\citep{Zevin:2020gbd,Cheng:2023ddt,  Riley:2023jep,colloms2025exploringevolutiongravitationalwaveemitters, Mastrogiovanni:2022ykr, Bouffanais:2019nrw, Bouffanais:2021wcr, Wong:2020ise, Mould:2022ccw, Plunkett:2025mjr}. Such approaches can also be considered reasonably flexible in terms of the underlying physical scenarios explored provided a diverse collection of simulated formation channels are used in the inference.

Normalizing flows~\citep[NFs,][]{rezende2016variationalinferencenormalizingflows,papamakarios2021normalizingflowsprobabilisticmodeling} are powerful generative models that can assist with both simulation-based inference~(SBI) and feature prediction, from PopSynth outputs. For example, \cite{colloms2025exploringevolutiongravitationalwaveemitters, Plunkett:2025mjr} use NFs for estimating the conditional distribution of CBC parameters for each synthetic population given values of the astrophysical initial conditions that characterize the simulated pathways. They perform simulation-based inference~(SBI) by efficiently interpolating the population likelihood over a grid of astrophysical initial conditions to constrain their posterior distribution given observational data from GW catalogs. On the other hand, NF emulators can also be used for data augmentation to predict population-level features and correlations in underrepresented regions of the CBC parameter space, where very few systems are synthesized from a particular channel, without having to simulate a prohibitively large ensemble of binaries~(as we demonstrate later on). However, like other deep learning emulators, the predictions of normalizing flows are susceptible to epistemic~(model) and aleatoric~(data) uncertainties~\citep{NEMANI2023110796,he2025surveyuncertaintyquantificationmethods}, which can, in principle, bias the inferred astrophysical conclusions.

For flow-based emulators, model uncertainties may arise either from inefficiencies in parameter learning, sub-optimal architectures, and out of distribution samples in the observed data, where as data uncertainties may result from the inherent sparsity and randomness of the training set which may be caused by particular formation channels being highly inefficient in certain regions of the CBC parameter space or by numerical approximation used by the simulators internally to speed up computation~\citep{NEMANI2023110796, he2025surveyuncertaintyquantificationmethods,andrews2024posydonversion2population}. While biases associated with ignoring model uncertainties may, in principle, be reduced by generating more training data, costly simulators can prohibit such endeavors. On the other hand, astrophysical bottlenecks leading to sparse and noisy datasets are difficult to mitigate against whilst simulating a tractable number of binaries. In other words, for SBI with state-of-the-art (costly) simulators or to correctly characterize their theoretical predictions for rarely synthesizable sub-populations such as highly unequal mass neutron Star black hole binaries~(NSBHs), BBHs with a component in the $3-5M_{\odot}$ mass range~\citep{LIGOScientific:2020zkf, Zevin:2020gma, LIGOScientific:2024elc}, CBCs emerging from stellar binaries at very high metallicities~\citep{metallicity}, and synthetic populations corresponding to large natal kicks etc., it is necessary to quantify and marginalize over emulator uncertainties for unbiased astrophysical inference. 

In this paper, we implement the Bayesian normalizing flow, a framework for emulating conditional and unconditional distribution functions equipped with uncertainty quantification and model comparison among architectures. By constructing the transformation layers of an NF from Bayesian neural networks~(BNNs) and exploiting the natural likelihood function associated with any density estimator model, such as an NF, we sample the posterior distribution of flow parameters given training data and suitable a priori assumptions. We use Hamiltonian Monte Carlo methods\citep[HMC, ][]{betancourt2018conceptualintroductionhamiltonianmonte} such as the No-U-Turn-Sampler~\citep[NUTS, ][]{Hoffman:2011ukg} to constrain the exact posterior distribution of flow parameters given training data, prior information, and the choice of model architecture. 

Using the posterior samples, we predict the emulated population as a Bayesian credible interval, rather than a deterministic density function, which is expected to contain the true population-level distribution of synthesized CBC parameters with a certain posterior probability, quantifying uncertainties in both density estimation~(necessary for SBI), and sample generation~(required for data augmentation and feature prediction). Using population synthesis simulations of BBHs formed through common envelope~(CE) evolution of isolated stellar binaries~\citep{Bavera_2020}, we show that the inferred credible intervals correctly quantify epistemic and aleatoric uncertainties in flow predictions by computing calibration curves, and demonstrate that they outperform an ensemble of flows trained on alternate realizations of sparse (thinned) datasets or on the same dataset with different parameter initializations. We further compute the Bayesian information criterion for different flow architectures to enable model comparison and thereby reduce the epistemic uncertainties that arise from sub-optimal architecture choices. We delineate how to marginalize the quantified uncertainties and perform model comparison among flow architectures for bias-free SBI and feature prediction using PopSynth simulations, as well as sketch additional applications in GW astronomy, whose implementations are left as futrue explorations. 


The rest of the paper is organized as follows. In Sec.~\ref{sec:motivation}, we highlight the main motivations for developing this framework, compare with previous studies in Sec.~\ref{sec:previous}, and describe the PopSynth datasets used for our demonstrations in Sec.~\ref{sec:data}. We discuss in Sec.~\ref{sec:methods:flows}, NF emulators for PopSynth simulations and demonstrate applications for data-augmentation and SBI. In Sec.~\ref{sec:methods:bnf} our detailed implementation of the Bayesian normalizing flow, its training, and calibration. We document our code and outline the applications of our uncertainty quantification framework in Secs.~\ref{sec:code}, \ref{sec:application}. In Sec.~\ref{sec:results}, we demonstrate the accuracy and interpretability of the quantified uncertainties for a fiducial BBH PopSynth dataset while also highlighting the role played by uncertainty quantification in data augmentation. We conclude in Sec.~\ref{sec:discussion} with a summary of the main developments presented and the follow-up investigations planned for the near future.

\section{Motivation: the necessity of UQ for high-fidelity simulators}
\label{sec:motivation}
The scope and reliability of SBI and feature prediction are determined by the fidelity of the simulator and its underlying physical prescriptions. Compared to its earlier versions~\citep{Bavera_2020}, which were used to model some isolated BBH formation scenarios in the training sets of \cite{colloms2025exploringevolutiongravitationalwaveemitters}, the \textsc{POSYDON} framework~\citep{Fragos_2023,andrews2024posydonversion2population} introduced significant advances in the treatment of binary stellar evolution, greatly expanding the astrophysical scope of SBI from GWs. \textsc{POSYDON} provides one of the most accurate descriptions of isolated CBC formation available in the literature by relying on extensive grids of binary evolution models that track stellar structure for both components using MESA~\citep{Paxton2011,Paxton2013, Paxton2015, Paxton2018, Paxton2019} from zero-age main sequence to compact object formation, combined with advanced interpolation methods~\citep{Fragos_2023}. In \textsc{POSYDONv2}, further improvements are achieved by accounting for evolution across a cosmological range of metallicities, reverse mass transfer, spin-orbit misalignment and binary disruption due to supernova kicks, and the possibility of stellar mergers~\citep{andrews2024posydonversion2population}, etc. A significant improvement in scalability, while retaining accuracy, is also achieved through the implementation of efficient machine learning algorithms. Altogether, \textsc{POSYDONv2} is emerging as a tracktable PopSynth simulator for isolated CBC formation (and many other astrophysical systems of interest) which boasts one of the highest physics fidelities to date. 

An immediate extension of existing astrophysical SBI from GW catalogs, facilitated by \textsc{POSYDON}, would be the derivation of constraints on both previously studied and additional astrophysical parameters—such as common envelop ejection efficiency, the slope of the initial mass function, supernova kick magnitudes, and parameters characterizing the cosmic star formation history—all based on accurate and reliable astrophysical modeling. Furthermore, synthesized binaries can be used to predict potentially new features in the population distribution of CBCs, should they arise from the improvements and generalizations of the underlying physical prescriptions implemented in binary modeling. The predicted features could, in principle, be modeled phenomenologically and hence constrained from data.

Therefore, given the high computational cost of running a detailed binary simulation, \textsc{POSYDONv2} can be assisted by NF emulators not only for SBI but also for feature prediction. For the latter, NFs can be used to augment the set of synthesized systems, which amounts to boosting sample statistics for a particular population, by learning from many smaller simulation sets at nearby initial condition grid points. In other words, instead of having to simulate millions of binaries for a single intitial condition of interest to properly sample under-represented regions of the CBC parameter space, many small simulation sets on a grid of nearby initial condition values~(that are paralelizable as individual POSYDON runs), can be used to train an NF emulator, which can then generate a large reliable sample of binaries for the particular initial conditions of interest. This can facilitate scalable and accurate population feature-prediction for rarely synthesizable CBCs without having to simulate a prohibitively large number of systems. Relevant examples include extreme mass ratio systems with high chirp mass, and populations of systems at high ($\geq Z_{\odot}$) metallicities~\citep{metallicity} or one that assumes large natal kick magnitudes~\citep{kicks}, tracktable \texttt{POSYDONv2} runs corresponding to which can be augmented by NF emulators to enable feature prediction with an accuracy is comparable to that of a much larger simulation. We show this later on for an existing dataset simulated with earlier versions of POSYDON, as proof of concept.

In other words, PopSynth emulators trained on a diverse collection of \textsc{POSYDONv2} runs promise novel and reliable astrophysical insights from growing CBC catalogs. However, generating arbitrarily many simulated binaries per initial condition grid point and using them to train emulators are both prohibitive, given the high computational cost of a single PopSynth run and that of NFs attempting to learn from billions or more datapoints, respectively. The latter might necessitate further downsampling of CBCs per population run in the SBI and/or feature-prediction training set. While NFs can facilitate data augmentation for a particular simulation grid point by learning from others (as we shall show later on), the predicted populations and density evaluations will be uncertain upon re-training with alternate realizations of down-sampled data or alternate initializations of flow parameters. The inherent randomness in \textsc{POSYDON} outputs due to numerical interpolation schemes used internally to speed up population synthesis, and sparsity in certain regions of CBC parameter space due to astrophysical bottlenecks, can be expected to further contribute to the variance in flow predictions for downsampled datasets. Emulating \textsc{POSYDON} runs for SBI and feature prediction, therefore, necessitates robust uncertainty quantification and marginalization to ensure reliable astrophysical explorations.
\section{Previous Work}
\label{sec:previous}
The implementation of Bayesian normalizing flows for conditional and unconditional density estimation, equipped with exact posterior sampling using HMC techniques that is presented here, and their application in the context of PoPSynth emulation, for SBI and data augmentation, and model comparison among flow architectures, is novel to the best of our knowledge. However, previous studies have developed UQ methods in an attempt to mitigate against the biases arising from uncertainties in flow-based deep learning models, which we discuss as follows.

In a previous study, \cite{Plunkett:2025mjr} discuss the effects of flow uncertainties in a very similar problem to ours, namely a simulation-based probe of Population III stars using next generation GW detectors. They train an ensemble of flows and average over the flow parameters from each training run, which, while capable of quantifying uncertainties, are, as they note, not strictly a Bayesian representation of model uncertainties~\citep[see also, ][ for an implementation of a Monte Carlo dropout-based ensemble network approach on NFs, ]{berry2023normalizingflowensemblesrich}. \cite{Plunkett:2025mjr} argue that their method outperforms approximate Bayesian techniques such as Fisher matrix analyses and variational inference, which are susceptible to exploring only single posterior modes, leading to poor uncertainty representation. However, their method can become intractable for a dense multi-dimensional grid of simulation initial conditions, or in the case of sparse and noisy training sets. 

Similarly, \cite{Ruhe:2022ddi} have used NFs to parametrize the population distribution of observed BBHs and inferred the population likelihood as a function of flow parameters. They have maximized said likelihood using gradient descent and explored uncertainties about the optimal flow parameters. As relevant to their study, they have used unconditional density estimators whose parameters they have inferred in the Bayesian sense from a likelihood that is specific to the problem of GW population inference. They have not sampled the posterior distribution of NF parameters given training data in the context of generic conditional and unconditional density estimation~(as needed for PopSynth emulation) nor explored the calibration and data augmentation impacts of the quantified uncertainties.

On the other hand, \cite{delaunoy2024} have investigated SBI with Bayesian neural networks~(BNNs) for low-budget ($O(10)$) simulations. They develop a functional prior on the flow parameters, the Bayesian model-average corresponding to which is well-calibrated by construction. Using mean field variational inference to train their BNNs, they demonstrate that the quantified uncertainties are better calibrated than the use of other priors on flow parameters. Again, they do not sample the exact posterior distribution for flow parameters.

In a different study, \cite{Bieringer:2024nbc} attempted Bayesian uncertainty quantification~(UQ) for unsupervised~(unconditional) density estimation using continuous normalizing flows~\citep{chen2018continuoustimeflowsefficientinference, grathwohl2018ffjordfreeformcontinuousdynamics}. They have either used approximate loss functions~\citep{lipman2023flowmatchinggenerativemodeling} instead of the exact log-likelihood or highly fine-tuned sampling methods~\citep{Bieringer:2023kxj} that have been subject to limited testing, to conduct their inference. They further restrict to a few~(O(50)) posterior samples for constructing calibration statistics and do not outline any model-comparison framework. Furthermore, since they restrict to only unconditional density estimators, generalizations of their method to SBI and data augmentation in the context of PopSynth emulators and its performance on test sets excluded from training remain unexplored. 

Instead, the method proposed here uses more robust, stable, and rigorously tested sampling techniques based on HMC methods, exact log-likelihoods, and calibration statistics constructed from thousands of posterior samples, which ensure stability and convergence. We sample multiple HMC chains parallely on separate GPUs to avoid exploring only one of multiple posterior modes. We further show that our method outperforms uncertainty estimation using NF ensembles for sparse datasets. While we use masked autoregressive flows~\citep{papamakarios2018maskedautoregressiveflowdensity} for the proof-of-concept demonstration presented here, this method is generalizable to continuous flows, which are expected to enable much faster convergence for HMC sampling. As we show later on, for advanced continuous flow algorithms~\citep[such as FFJORD,][]{grathwohl2018ffjordfreeformcontinuousdynamics} with scalable log density evaluations, the complexity required to emulate realistic CBC PopSynth simulations is also well within the scope of modern HMC samplers. Hence, a PopSynth emulator based on Bayesian continuous flows would not necessitate highly fine-tuned samplers such as the ones proposed by~\cite{Bieringer:2023kxj}. This endeavor is left as a future exploration.

\section{Population Synthesis datasets}
\label{sec:data}
To demonstrate the performance and accuracy of our method, we use population synthesis simulations of merging BBHs formed through the common envelope evolution of isolated stellar binaries~\citep{Bavera_2020}. The formation of these systems was modeled using an earlier version of the \textsc{POSYDON} framework, which combines various stages of binary evolution simulated by the rapid population synthesis code \textsc{COSMIC}~\citep{Breivik:2019lmt} with the detailed stellar structure calculations performed by \textsc{MESA}~\citep{Paxton2011,Paxton2013,Paxton2015,Paxton2018,Paxton2019}, and is described in ~\citep{Fragos_2023}. Each population in the simulated dataset is characterized by two initial conditions, namely the common envelop ejection efficiency $\alpha$ and the BH natal spin $\chi_b$. The simulations were carried out over a grid of initial conditions spanned by $\alpha \in \{0.2, 0.5, 1.0, 2.0, 5.0\}$ and $\chi_b \in \{0, 0.1, 0.2, 0.5\}$. We chose the population corresponding to $\alpha=1.0,\chi_b=0.1$ to be our test set by excluding it from both training, validation, and Bayesian inference.
\section{Normalizing Flow Emulators for population synthesis simulations}
\label{sec:methods:flows}

We start by noting that PopSynth simulations generate an ensemble of systems with uniquely distributed parameters ($\vec{\theta}$) given a particular set of initial conditions and/or other quantities ($\vec{\lambda}$) that characterize the evolutionary track being simulated. Hence, emulating PopSynth essentially amounts to constructing accurate estimators for the conditional distribution function:

\begin{equation}
    p(\vec{\theta}|\vec{\lambda})
\end{equation}
Normalizing flows~(NFs, \cite{rezende2016variationalinferencenormalizingflows, papamakarios2021normalizingflowsprobabilisticmodeling}) have emerged as scalable and highly expressive generative models that can learn the underlying distribution of data and thereby generate new datapoints similar to the training set. They also facilitate tractable density evaluation at intermediate points in the parameter space that are not part of the training set. In other words, NFs are capable of accurate conditional and unconditional density estimation from labeled and unlabeled datasets, respectively. For labeled datasets such as PopSynth outputs, an NF model estimates the target conditional density as a transformed base-distribution, wherein the transformation functions are usually approximated by deep neural networks~(DNNs):
\begin{eqnarray}
    \hat{p}(\vec{\theta}|\vec{\lambda},\vec{\omega}) = p_{u}(\vec{u}=f_{NN}(\vec{\theta},\vec{\lambda},\vec{\omega}))\left|\frac{\partial\vec{u}}{
    \partial \vec{\theta}}\right|.
\end{eqnarray}
Here $\vec{\omega}$ are the parameters of the DNN, $p_u$ is a base distribution comprised of i.i.d. standard Gaussian, and $f_{NN}$ is a sequence of differentiable and invertible transformations. Once the model is trained to find an optimal $\vec{\omega} = \vec{\omega}_0$, samples can be drawn from an emulated population in between the training point $(\vec{\lambda} = \vec{\lambda}_{new})$ by first drawing $\vec{u}_i\sim p_u$ and then computing $\vec{\theta}_i=f^{-1}_{NN}(\vec{u}_i,\vec{\lambda}_{new}, \vec{\omega}_0)$. Data augmentation for training points is also possible for boosting sample-statistics in under-represented subpopulations for improving feature prediction. Similarly, the conditional density becomes tractable as a numerical function suitable for repeated evaluations, also in between training points $(\vec{\theta} = \vec{\theta}_{new}, \vec{\lambda} = \vec{\lambda}_{new})$, which is required to implement SBI from observed catalogs~\citep{colloms2025exploringevolutiongravitationalwaveemitters}. 

Various architectures have been proposed in literature for constructing $f_{NN}$, each with its own merits and demerits. Broadly, the transformation functions can be classified into two categories: discrete layers of finite depth, such as masked autoregressive flows~\citep[MAF, ][]{papamakarios2018maskedautoregressiveflowdensity}, and continuous time transforms, which are essentially diffusion models that asymptotically approach the target density~\citep {chen2018continuoustimeflowsefficientinference, grathwohl2018ffjordfreeformcontinuousdynamics}. While various sub-categories under these two classes can be advantageous to use for specific problems, continuous normalizing flow~(CNF) algorithms such as FFJORD~\citep{grathwohl2018ffjordfreeformcontinuousdynamics} are, in general, more scalable and require DNNs of much lower complexity to obtain the same expressivity as a discrete flow like MAF. We shall show this in the context of our PopSynth datasets while noting that the lower complexity makes CNFs ideal for the Bayesian framework proposed for UQ.

\begin{figure*}[htt]
\begin{center}
\includegraphics[width=0.46\textwidth]{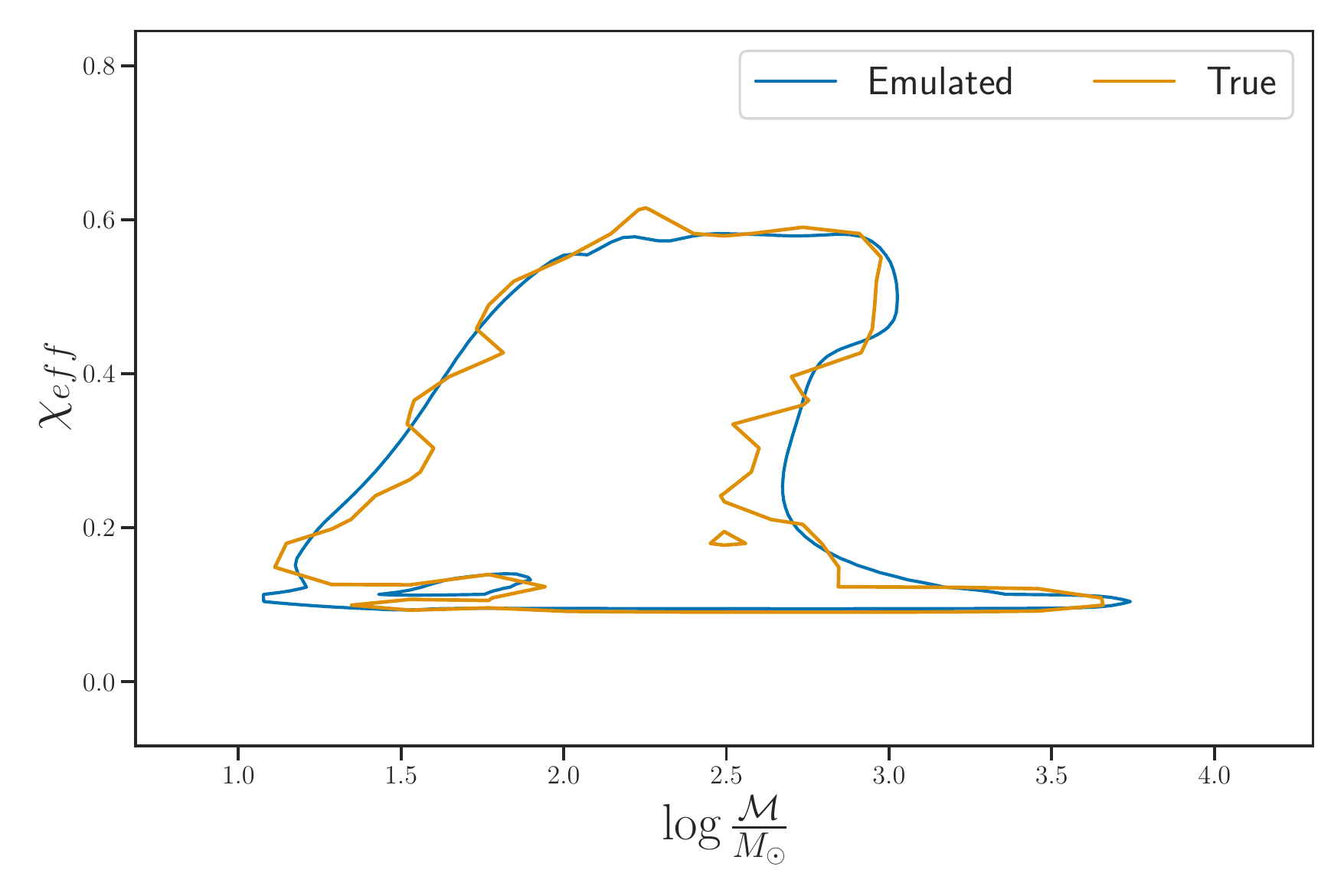}
\includegraphics[width=0.4\textwidth]{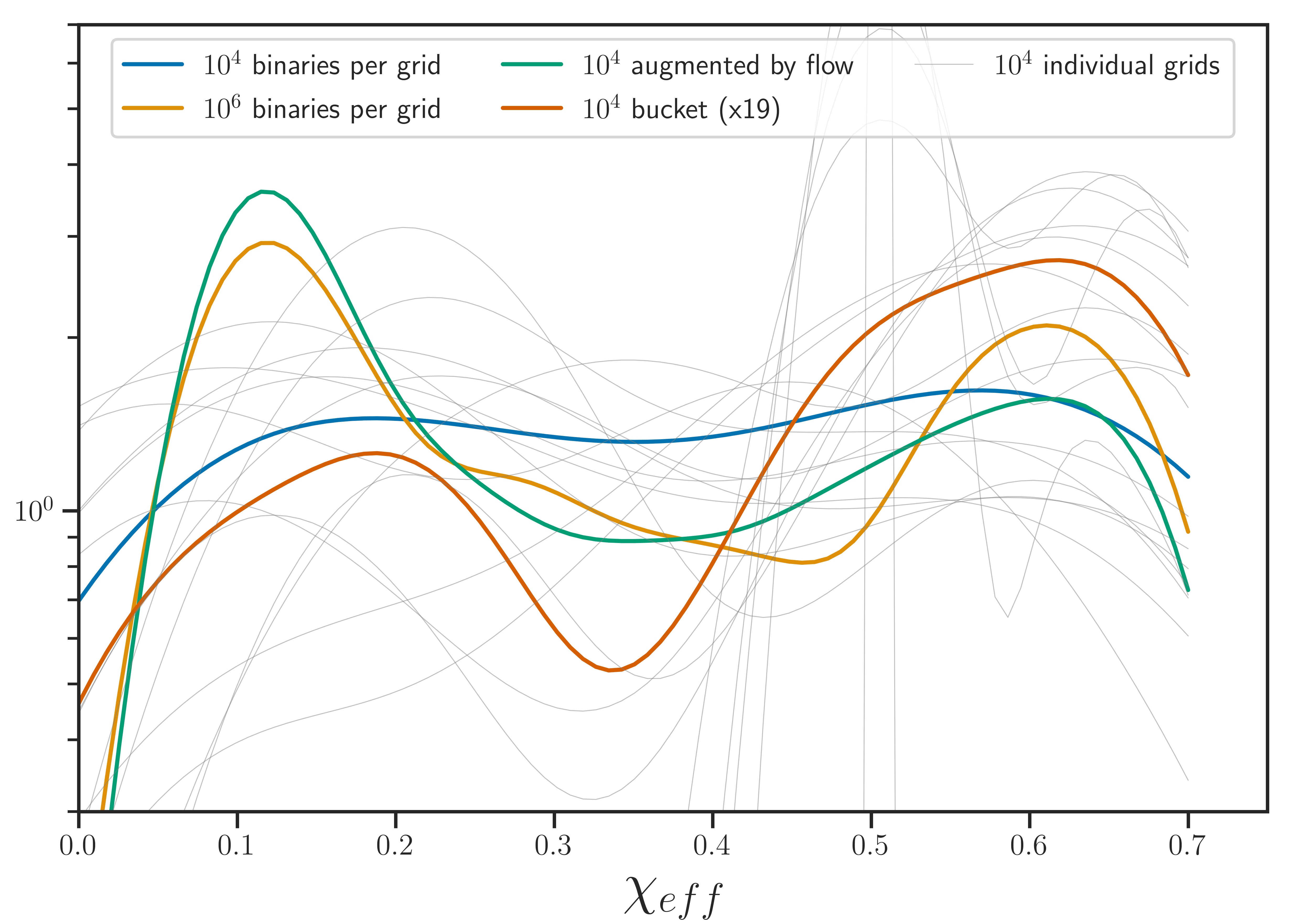}
\caption{\label{fig:amp} Density estimation~(left) for SBI, and data augmentation (right) for feature-prediction with a MAF-based PopSynth emulators. For the right pannel we focus on under-represented regions of parameter space. Each curve represents the effective spin distribution of BBHs in the test set that belong to an unequal mass ratio and high chirp mass sub-population.}
\end{center}
\end{figure*}

For any given transformation architecture, we can train the NF model by minimizing some loss function with respect to $\vec{\omega}$, such that the optimal $\vec{\omega}$ corresponds to $p$ and $\hat{p}$ being maximally similar~\citep{rezende2016variationalinferencenormalizingflows}. As a standard metric, the \textit{expected} KL divergence (averaged over the list of simulations on which we want to train) is chosen. Assuming the training simulations where carried out for gridpoints: $\{(\vec{\lambda})\sim p_{\lambda}(\vec{\lambda})\}$, the training dataset:  $\vec{d}_{train}=\{(\vec{\theta}_i, \vec{\lambda}_i)\sim p(\vec{\theta},\vec{\lambda})\}_{i=1}^{N}$ can be thought of as a set of i.i.d. samples drawn from the joint distribution $p(\vec{\theta},\vec{\lambda}) = p(\vec{\theta}| \vec{\lambda})p_{\lambda}(\vec{\lambda})$. Under these considerations, the expected KL divergence can be shown to equal the following sum over training datapoints (up to an additive constant):
\begin{eqnarray}
    D_{KL}(\vec{\omega},\vec{d}_{train}) = -\frac{1}{N}\sum_{(\vec{\theta}_i, \vec{\lambda}_i)\in \vec{d}_{train}} \log\hat{p}(\vec{\theta}_i|\vec{\lambda}_i,\vec{\omega})+const \label{eq:KLLL}
\end{eqnarray}
The best-fit model can be obtained by minimizing this loss function with respect to the model parameters $\omega$ using some gradient descent algorithm such as \textsc{Adam}\citep{kingma2017adammethodstochasticoptimization}. The optimal parameters corresponding to the trained model can be represented as functions of the training dataset: $\vec{w}_{0}(\vec{d}_{train})$. The trained emulator in this notation is therefore:
\begin{eqnarray}
    \bar{\hat{p}}(\vec{\theta}|\vec{\lambda}, \vec{d}_{train})= \hat{p}(\vec{\theta}|\vec{\lambda}, \vec{\omega}_0(\vec{d}_{train})),  
\end{eqnarray}
which can be used for tractable density evaluation, data generation, and augmentation. In other words, the trained emulator can, in principle, facilitate SBI and feature prediction for state-of-the-art simulators such as \textsc{POSYDONv2}, both of which are demonstrated in figure~\ref{fig:amp}, for a BBH PopSynth dataset that uses earlier versions of \textsc{POSYDON} to simulate common-envelope evolution of isolated stellar binaries~\citep{Bavera_2020}.

While \cite{colloms2025exploringevolutiongravitationalwaveemitters} have extensively discussed SBI using MLE flows, we show here another astrophysical application of PopSynth emulation. Say one is interested in predicting effective spin distributions of a sub-population of BBHs formed through CE, astrophysical parameters same as our test set, that have unequal mass ratio~($<0.4$) and high chirp mass~($>7M_{\odot}$). The CE channel struggles to populate this region of the parameter space~\citep{Zevin:2020gma}. If as few as $O(10^4)$ binaries are simulated for this grid point, only $O(10)$ systems fall in this region of parameter space, which are not sufficient for accurate feature prediction. 

On the other hand, an NF trained on multiple population runs, each with $10^4$ binaries corresponding to different grid points can be used to accurately amplify the dataset of interest, even if it is excluded from training, without having to run a larger PopSynth simulation. In Fig.~\ref{fig:amp}, the mentioned spin distribution is shown for the small simulation set, the amplified set obtained using an NF trained on all available simulations except the one being amplified, and a much larger PopSynth rerun~($10^6$ binaries) which is treated as the ground truth. It can be seen that the emulator predicts the distribution with reasonable accuracy despite being trained on a collection of highly downsampled populations. In other words, NFs can be used to predict population features of rarely observed systems without having to simulate a very large number of binaries, which can be prohibitive. Similar advantages are expected to arise in the context of feature prediction with \texttt{POSYDONv2} for CBCs formed from isolated stellar binaries with high metallicities or high natal kicks~\citep{metallicity, kicks}.

\subsection{Uncertainties in flow predictions}
\label{sec:methods:uncert}
The KL divergence as a function of $\vec{\omega}$ can be complicated with multiple peaks, troughs, and saddle points, leading to gradient-based minimization algorithms finding slightly different answers for the best-fit estimator every time the model is re-trained. Loss function morphology can also change significantly between alternate realizations of thinned and noisy training data.  In other words, if we train the same model on the same dataset twice, or on different realizations of a noisy dataset, we won't necessarily get the exact same answer for the predicted population. It is important to quantify and marginalize over this variance in flow predictions for accurate SBI and feature prediction, particularly in the cases of sparse and noisy training datasets and sub-optimal architecture choices. Hence, instead of a best-fit emulator that deterministically yields a single distribution, it is desirable to predict a range of density functions that are likely to contain the true population with a certain posterior probability given training data, model assumptions, and a priori expectation on DNN parameters. Before we show how that can be achieved, it is worthwhile to discuss the types and sources of these uncertainties that we wish to quantify in a bit more detail.

As mentioned, uncertainty in flow predictions, or DNN models in general, can arise in a variety of scenarios and hence can be classified based on their source and nature, which are often directly correlated~\citep{NEMANI2023110796,he2025surveyuncertaintyquantificationmethods,H_llermeier_2021}. Inefficiencies in parameter learning, sub-optimal architecture choices, and out of distribution inference samples ($p_{\lambda}(\vec{\lambda}_{new})\rightarrow0$), can give rise to stochastic and systemic uncertainties in the optimal model predictions, which can be biased and varying between training re-runs. These \textit{epistemic}/model uncertainties are, in principle, thought to be reducible given additional training data. In the absence of additional data, these uncertainties can be quantified using various approaches such as Bayesian neural networks~\citep[BNNs, e.g.,][]{arbel2023primerbayesianneuralnetworks}, ensemble networks~\citep[e.g.,][]{Ganaie_2022}, and Monte Carlo dropout~\citep{gal2016dropoutbayesianapproximationrepresenting} etc., which have been studied less commonly for frameworks in which an NF is the underlying NN model whose uncertainties are being quantified.

On the other hand, uncertainty can also arise from the inherent randomness, sparsity, and class-confusion present in the dataset, which is irreducible even with the addition of more training samples~\citep{he2025surveyuncertaintyquantificationmethods, NEMANI2023110796, H_llermeier_2021}. Key examples of these \textit{aleatoric}/data uncertainties in the PopSynth of merging compact binaries are the sparsity of certain regions of parameter space, such as extreme-mass ratio binaries with ``mass-gap" compact objects~\citep{Zevin:2020gma,LIGOScientific:2020zkf} or CBCs emerging from high metallicity stellar binaries, where the merger rate is expected to be low, degenerate population features shared by different values of $\vec{\lambda}$, and the inherent randomness resulting from numerical interpolation schemes used by simulators like \texttt{POSYDONv2}. They would become particularly relevant when \texttt{POSYDONv2} is used for SBI from catalogs including mass-gap NSBH-like systems, of which two have already been observed~\citep{LIGOScientific:2020zkf, LIGOScientific:2024elc}, and for making tractable theoretical predictions for similar or other rarely synthesizable sub-populations. Quantifying data uncertainties in deterministic DNN predictors can be achieved through Bayesian inference~\citep{arbel2023primerbayesianneuralnetworks} and parametric or non-parametric modeling of the distribution of noise in the data, as well as the use of deep probabilistic models such as secondary NFs and generative adversarial networks~\citep{H_llermeier_2021, whang2021composingnormalizingflowsinverse, dirmeier2023uncertaintyquantificationoutofdistributiondetection}. Again, such approaches have not been widely used to quantify the uncertainties of NF parameters, which is relevant in scenarios where the NF itself is the underlying DNN predictor.

In the following, we present a new implementation of the Bayesian NF, a generative probabilistic model constructed from BNNs, which can self-consistently quantify epistemic uncertainties that arise from inefficient training and aleatoric uncertainties that arise from the sparsity and stochasticity inherent to the dataset. No approximations for the log-likelihood function are utilized, and rigorous sampling methods are used for Bayesian inference. The resulting predictions for an emulated PopSynth simulation are hence marginalized over such uncertainties and also automatically enable Bayesian comparison among different NF models, thereby minimizing the epistemic uncertainties that arise from choosing sub-optimal architectures and families of networks.
\section{Uncertainty Quantification and Model Comparison: Bayesian Normalizing Flows}
\label{sec:methods:bnf}
To quantify these epistemic and aleatoric uncertainties in the NF model for thinned datasets, we propose the following. As can be seen from Eqn.~\eqref{eq:KLLL}, The KL divergence is proportional to the negative log-likelihood of the model parameters given the training dataset (up to an additive constant):
\begin{equation}
    \log p(\vec{d}_{train}|\vec{\omega}) = \sum_{(\vec{\theta}_i, \vec{\lambda}_i)\in \vec{d}_{ttain}} \log\hat{p}(\vec{\theta}_i,\vec{\lambda}_i|\vec{\omega}) = \sum_{(\vec{\theta}_i, \vec{\lambda}_i)\in \vec{d}_{train}} \log\hat{p}(\vec{\theta}|\vec{\lambda},\vec{\omega}) + const. \label{eq:loglike}
\end{equation}
where $\log\hat{p}(\vec{\theta},\vec{\lambda}|\vec{\omega}) = \log\hat{p}(\vec{\theta}|\vec{\lambda},\vec{\omega})+\log p_{\lambda}(\vec{\lambda})$. Note that \emph{this is the exact log-likelihood of the parameters characterizing a probabilistic (NF) model, and no a priori assumption about the distribution of noise in the dataset is necessary to construct its functional form}. The training method in Sec.~\ref{sec:methods:flows} above, therefore, amounts to finding a maximum likelihood density estimate~(MLE). To quantify uncertainties, we instead propose to sample the posterior distribution of the model parameters using some suitably chosen prior and the exact log-likelihood:
\begin{equation}
    \log p(\vec{\omega}|\vec{d}_{train}) = \log p(\vec{d}_{train}|\vec{\omega})+\log p_{\omega}(\vec{\omega})+const \label{eq:logpost}
\end{equation}
where $p_{\omega}(\vec{\omega})$ is the prior distribution of the model parameters, which can be chosen to have support only nearby $\vec{\omega}_0$ to speed up convergence in posterior exploration. Stochastically sampling the posterior using robust and rigorously established Hamiltonian Monte Carlo~(HMC, \cite{betancourt2018conceptualintroductionhamiltonianmonte}) techniques such as the No-U-Turn Sampler~\citep[NUTS,][]{Hoffman:2011ukg}, one can obtain the posterior draws for the flow parameters: $S_{\vec{\omega}}=\{\vec{\omega}_i \sim p(\vec{\omega}|\vec{d}_{train})\}_{i=1}^M$. These can in turn be used to evaluate a trained density estimator marginalized over model uncertainties:
\begin{equation}
    \bar{\hat{p}}_m(\vec{\theta}|\vec{\lambda},\vec{d}_{train}) = \int \hat{p}(\vec{\theta}|\vec{\lambda},\vec{\omega}) p(\vec{\omega}|\vec{d}_{train}) = \frac{1}{M}\sum_{\vec{\omega_i}\in S_{\vec{\omega}}} \hat{p}(\vec{\theta}|\vec{\lambda},\vec{\omega}_i)
\end{equation}
where $_m$ stands for marginalized. This amounts to constructing and training a normalizing flow whose transformations comprise a BNN. Alternatively, the predicted population can be represented as a credible interval instead of a single distribution, which is expected to contain the true population with a certain posterior probability, given training data and prior expectations on the model parameters. The direct application of this framework to SBI and data augmentation is outlined in Sec.~\ref{sec:application}.

While the posterior samples only account for model uncertainties within a fixed architecture, this framework can also be used to perform model selection between architectures. Once the posterior distributions of two NF models with different architectures have been sampled, one can compute the Bayesian information criterion~(BIC) for both of them and select the one with the lowest BIC as the preferred architecture with the least amount of overfitting. In our notation, the BIC takes the following form:
\begin{equation}
    \mathrm{BIC} = n_{\omega}\ln{N} - \ln{p_{max}(\vec{d}_{train}|\vec{\omega})},
\end{equation}
where $n_{\omega}$ is the complexity of the model, $N$ is the number of points in the training set, and $p_{max}(\vec{d}_{train}|\vec{\omega})$ is the maximum likelihood value reached during posterior sampling. 

The full algorithm for uncertainty quantification and model comparison is sketched in \ref{alg:BNF}. Note that unlike regular Bayesian networks such as the ones used in regression, which needs restrictive assumptions (such as Gaussianity of uncertainties) to construct the likelihood from the loss function~\citep{arbel2023primerbayesianneuralnetworks}, the Bayesian NFs proposed here are equipped with the natural likelihood function of the distribution estimator which requires no implicit assumption about the distribution of uncertain model parameters.

On the other hand, NFs have been used in literature to model the likelihood of other Bayesian networks~\citep{dirmeier2023uncertaintyquantificationoutofdistributiondetection, whang2021composingnormalizingflowsinverse}. In such studies, the parameters of the NF itself are trained using the MLE method with only the parameters of the underlying network inferred in the Bayesian sense. In such studies, the NF merely serves as a tool for quantifying the uncertainties of the underlying DNN model, with the uncertainties of the flow itself ignored as a second-order effect. However, for emulating PopSynth, the NF itself is the underlying generative model, which necessitates UQ of flow parameters through the proposed Bayesian NF methodology. In addition, the proposed framework can be used to quantify uncertainties self-consistently in NF-based likelihood-free inference frameworks~(see appendix~\ref{sec:appendix:NLE}), as well as the mentioned attempts at uncertainty quantification of other architectures that rely on MLE-trained NFs, should the latter be applied to scenarios that necessitate the marginalization over second-order NF uncertainties.

\subsection{Prior choices, Implementation and Scalability}
Choosing the prior distribution on model parameters correctly is crucial for convergent and scalable Bayesian inference, particularly for models of such high complexity. To fully explore the likelihood function, highly restrictive priors should be avoided. However, broad priors such as standard Gaussians or broad uniform distributions might lead to slow convergence for highly complex probabilistic models such as discrete NFs that are usually characterized by $O(10^5-10^6)$ parameters. For highly complex models such as MAFs, it is instead desirable to choose a prior that is only supportive near the MLE parameters and yet broad enough for likelihood-driven inference. Furthermore, instead of fixing the width of the prior, it can also be allowed to vary within a certain range. We chose the following simple choice for the prior distribution for this proof-of-concept demonstration.

We start by choosing a maximal scale $\vec{\sigma} = \{\sigma_0\}_{i=1}^{n_{\omega}}$, which amounts to assuming a priori that the highest likelihood model parameters are within a factor of at most $\sigma_0$ of $\vec{\omega}_0$. we then define the standardized model parameters $\vec{\omega}_{st} \underset{i.i.d}{\sim} \mathcal{U}(-1,1)$ which we assume to be distributed uniformly within $(-1,1)$. we then rescale these to become numbers that lie within a factor of $\vec{\sigma}$ of $\vec{\omega}_0$ and hence construct the prior on $\vec{\omega}=\vec{\omega}_0(1+\vec{\sigma}\vec{\omega}_{st})$ as an affine transformed uniform distribution. The prior scale $\vec{\sigma}$ can, in principle, be sampled as well from a uniform or half-normal distribution, but for simplicity, we fix it to a value that allows for a reasonably broad prior and likelihood-driven inference. As we demonstrate shortly, calibration statistics can be constructed for choosing optimal values of $\sigma_0$.

While these priors might seem restrictive as compared to a standard Gaussian, they can be expected to suffice as long as the posterior is found to be more tightly constrained, away from the prior boundaries. We show that this is the case in the context of our population synthesis dataset and suggest increasing $\sigma_{0}$ for scenarios wherein the posterior is found to be railing against the prior. More sophisticated prior choices that are less restrictive and yet consistent with fast convergence are left as a future exploration.

Even with these prior choices, the Bayesian NFs can quickly become intractable for high-dimensional distributions since discrete flows like MAF require highly complex networks to accurately capture all the features in the population. Bayesian inference can then amount to exploring (and storing) potentially thousands of copies of nearly a million-parameter model, which is expected to become prohibitive. As we show later on in appendix~\ref{sec:appendix:cnf}, CNFs can solve this problem by reducing the required model complexity for accurate density estimation by several orders of magnitude. Since the proposed Bayesian framework is agnostic of model architecture, it can be expected to work with CNFs in the same way as discrete flows, and thereby quantify uncertainties in the emulators of higher-dimensional population distributions. On the other hand, if reducing model complexity through the use of CNFs is not a viable alternative, or if complex embedding nets are necessary to reduce the dimensionality of $\vec{\lambda}$, approximate Bayesian inference can be implemented through a combination of stochastic variational inference and importance sampling which is outlined in appendix~\ref{sec:appendix:svi}.

We leave the full realization of both of these resolutions for the future and restrict to Bayesian MAFs that emulates a two-dimensional synthetic population, and a four-dimensional one, in this proof-of-concept demonstration. As mentioned, the high complexity of the MAF makes Bayesian inference with fully un-restricted priors, and large datasets, infeasible. This also results in BICs that are likely driven more by the change in complexity than that in the maximum likelihood achieved during sampling, regardless of overfitting, given the size of the datasets necessitated by tractable inference. We are, however, able to demonstrate that exact posterior sampling can still facilitate likelihood-driven inference that is more informative than the prior, well-calibrated UQ, and BICs that penalize overtly complex models. 
\subsection{Calibration of Uncertainties}
To appraise the accuracy of the estimated uncertainties and optimize the choice of $\sigma_0$, we construct metrics that compare the empirical coverage of a test set to the theoretical coverage predicted by Bayesian inference~\citep{Bieringer:2024nbc}. After drawing M posterior samples of flow parameters, we split them up into m random subsets: $S_{\vec{\omega}}=\bigcup_{i=1}^{m} S_{\vec{\omega}}^i$, each representative of a smaller Bayesian inference run. For every posterior sample in a particular subset, we generate a million binaries $\vec{\theta}\in S_{\vec{\theta}}^{ij}=\{\vec{\theta}_k\sim \hat{p}(\vec{\theta}|\vec{\lambda}_{test},\vec{\omega}_j)|\vec{\omega}_j\in S_{\vec{\omega}}^i\}_{k=1}^{10^6}$. we then construct $n_{Q}$ equal probability mass $\vec{\theta}$ bins~(quantile binning) from the test set and compute the count of predicted samples in each bin. We then compute credible intervals for the counts and estimate the empirical coverage as the fraction of times (out of m and averaged over bins) the credible interval encloses the actual count of test samples. The calibration curve is then constructed as the relationship between the empirical coverage and the nominal coverage~\citep[confidence level of the credible intervals, ][]{Bieringer:2024nbc}. Note, however, that for very large values of $\sigma_0$, if all of the walkers get stuck at a local maxima of the likelihood function, there will be biases resulting in low coverage. In other words, low coverage can be indicative of both under-estimation of uncertainties due to small $\sigma_0$ or biases due to very large $\sigma_0$, causing difficulty in convergence. This issue was not discussed in \cite{Bieringer:2024nbc}. 

 For comparison, we also compute calibration curves for an ensemble of flows. We re-train emulators using the MLE method on alternate realizations of downsampled datasets (of the same size as the ones used to train the Bayesian flow) and compute the empirical coverage of the test set with respect to the credible intervals yielded by the ensemble. We show that even for ensembles comprised of hundreds of emulators, the Bayesian method is better calibrated than with regard to the true distribution.
\begin{algorithm}[H]
\caption{Bayesian Normalizing Flow \label{alg:BNF}}
\begin{algorithmic}[1]
\Require $(N,d_{\theta}, d_{\lambda})$: data dimensions; $\vec{d}_{train}\in \mathbb{R}^{N \times (d_{\theta}+d_{\lambda})}$: training data
\Require $(\textsc{Adam}, \alpha, \beta_1, \beta_2)$: Optimizer and associated parameters \Comment{From \cite{kingma2017adammethodstochasticoptimization}}
\Require $(\textsc{NUTS}, \epsilon, M)$: Sampler and associated parameters \Comment{From \cite{Hoffman:2011ukg}}
\Require $F(\vec{\theta}, \vec{\lambda}, \vec{\omega})$: Flow model, computes $\hat{p}(\vec{\theta}| \vec{\lambda}, \vec{\omega})$
\Require $n_\omega$: model complexity, $\vec{\sigma} $: fixed prior scale for each model parameter
\Require $avg\in[\mathrm{True}, \mathrm{False}]$: whether to use the average or exact log-likelihood
\Ensure $S_{\vec{\omega}}, \text{BIC}$

\State $\vec{\omega}_{init}\sim \mathcal{N}(0, I)$; $\vec{\omega}_0 \gets \textsc{Adam}(\alpha, \beta_1, \beta_2, \textsc{Loss}, \vec{\omega}_{init})$
\State $S_{\vec{\omega}} \gets \textsc{NUTS}(\vec{\omega}_{init}, \epsilon, \textsc{LogPost}, M)$
\State $L_{max}\gets -\infty$
\For{$i=1$ to $M$}
  \State $L \gets \textsc{LogLikelihood}(S_{\vec{\omega}}[i])$;\quad\textbf{if} $L > L_{max}$ \textbf{then} $L_{max}\gets L$
\EndFor
\State $\text{BIC} \gets n_{\omega}\log N - L_{max}$

\Function{LogPost}{$\vec{\omega}$}
  \State \Return $\textsc{LogLikelihood}(\vec{\omega}, avg) + \textsc{LogPrior}(\vec{\omega})$ 
\EndFunction

\Function{LogPrior}{$\vec{\omega}$}
  \State $P \gets - \log{2} - \sum_{i=1}^{n_{\omega}} \log(\vec{\omega}_0[i]\vec{\sigma}[i])$
  \State \Return $P$ \quad \textbf{if} $\frac{\vec{\omega}-\vec{\omega}_0}{\vec{\sigma}\times\vec{\omega}_0 (\textsc{elementwise})}\in[-1,1]$ \quad \textbf{else} $-\infty$
\EndFunction

\Function{Loss}{$\vec{\omega}$}
  \State \Return $-\textsc{LogLikelihood}(\vec{\omega}, \mathrm{True})$
\EndFunction

\Function{LogLikelihood}{$\vec{\omega},avg$}
  \State $L \gets \sum_{i=1}^N \log f(\vec{d}_{train}[i,1{:}d_{\theta}],\vec{d}_{train}[i,d_{\theta}{:}], \vec{\omega})$
  \State \Return $L/N$ \textbf{if} $avg$ \textbf{else} $L$
\EndFunction
\end{algorithmic}
\end{algorithm}

\section{Application: Simulation-based inference and data augmentation with UQ}
\label{sec:application}
As shown by \cite{colloms2025exploringevolutiongravitationalwaveemitters}, an NF emulator can be used to construct the likelihood of the simulation initial conditions given data $\vec{d}_{o}$ from an observed population of systems, using Bayesian hierarchical inference. However, since they train their flows using the MLE method, their inference is always conditional on a fixed value of the flow parameters $\vec{\omega}= \vec{\omega}_0$. We instead propose to use the posterior samples of the flow-parameters to obtain empirical constraints on $\vec{\lambda}$ that are marginalized over flow uncertainties using the following posterior distribution:
\begin{equation}
    p(\vec{\lambda}|\vec{d}_o, \vec{d}_{train}) = \pi(\vec{\lambda})\int d\vec{\omega}p(\vec{d}_o|\vec{\lambda},\vec{\omega})p(\vec{\omega}|\vec{d}_{train}) 
\end{equation}
where $p(\vec{d}_o|\vec{\lambda},\vec{\omega})$ is essentially the LHS of Eq.~(1) of \cite{colloms2025exploringevolutiongravitationalwaveemitters}, for a particular value of $\vec{\omega}$, (which they choose to be the MLE corresponding to a single training run). To infer $\vec{\lambda}$, we propose to categorically select $\vec{\omega}\in S_{\vec{\omega}}$ for each draw of $\vec{\lambda}\sim\pi(\vec{\lambda})$ explored during sampling the latter's posterior distribution and use the flow corresponding to the selected $\vec{\omega}$ values for computing the likelihood $p(\vec{d}_o|\vec{\lambda},\vec{\omega})$.

Similarly, in the context of data augmentation for rarely synthesizable sub-spopulations, we propose the following. Instead of using the MLE flows trained on a grid of simulations to draw a large number $\vec{\theta}$ samples, we suggest generating an ensemble of augmented datasets, one corresponding to each posterior draw of the flow parameters: $\vec{d}_{A}(\vec{\lambda}_{test})=\{\vec{\theta}_{ij}\sim \hat{p}(\vec{\theta}|\vec{\lambda}_{test}, \vec{\omega}_j)|\vec{\omega}_j\in S_{\vec{\omega}}\}$. This dataset can be used to provide a credible interval of the predicted density function for the sub-population in question, which will contain the true prediction as well as that of a single MLE flow with a certain posterior probability. We show this in our demonstration.

\section{Code and Data}
\label{sec:code}
The code developed to implement this framework is publicly available in the form of the \textsc{Python} package \textsc{naz}(Normalizing flow Algorithms beyond Zero-variance training), which can be found at \url{https://github.com/AnaryaRay1/naz}. In addition to MAF and CNF, additional architectures such as neural spline coupling and spline autoregressive flows~\citep{durkan2019neuralsplineflows} are provided, with every model capable of both conditional (supervised) and unconditional (unsupervised) density estimation. The MAF is implemented in both a \textsc{Pytorch}+\textsc{Pyro}~\citep{paszke2017automatic, bingham2019pyro} and \textsc{Jax+Numpyro}~\citep{jax2018github, phan2019composable} backend with functionality provided for converting one to the other. The \textsc{Jax+Numpyro} backend has been found to be more suitable for Bayesian inference, while the \textsc{Pytorch}+\textsc{Pyro} one works better for standard (MLE) training. The other architectures are only available with the \textsc{Pytorch}+\textsc{Pyro} backend while the CNF additionally requires the \textsc{torchdyn}~\citep{poli2020torchdynneuraldifferentialequations} package for implementing the neural ode solvers needed by the FFJORD algorithm. \textsc{Jax+Numpyro} backends for these architectures will be available in future releases of \textsc{naz}. The training data used in the demonstrations is obtained from publicly available BBH PopSynth simulations released in zenodo by \cite{Bavera_2020}, details of which are described in the next section. The trained flows, posterior samples, and smaller training sets will soon be available in zenodo for reproducibility.

\section{Illustrative Results}
\label{sec:results}

NF emulators can facilitate both SBI and feature prediction from simulation sets described in Sec.~\ref{sec:data}, which are necessary for gaining astrophysical insights from observed CBC catalogs, now equipped with uncertainty quantification and model selection. For the training set, we include 10000 BBHs per grind point for the MLE method used to demonstrate feature prediction and heavily downsample to 20 binaries per grid point for Bayesian inference. Similar conclusions will emerge for less downsampling, but will take longer run-times and additional hardware resources to converge for the current MAF-based implementation. To demonstrate how Bayesian NFs can mitigate the biases resulting from un-marginalized flow variance, we also train an ensemble of 150 flows using the MLE method on many different realizations of the smaller (20 BBHs per grid point) and compare the corresponding density estimates with the posterior predictive emulator provided by Bayesian inference. We further show that the Bayesian uncertainty estimates are better calibrated than those obtained through an ensemble of MLE re-runs for such sparse datasets.

The default MAF architecture chosen for all illustrative results comprises 16 masked autoregressive layers, each of which contains 3 hidden layers spanned by 150 units. MLE training was carried with a learning rate decay of 0.5 and early stopping implemented by means of a validation set to minimize over-fitting. With batch sizes of 100, training is fairly efficient on NVIDIA-A100 GPUs with 40GB of memory. For Bayesian inference, however, 4 NVIDIA-H100 GPUs with 80GB of HB200 ram were necessary for running 4 HMC chains with 500 tuning and 1500 sampling steps. These requirements are expected to be reduced significantly once Bayesian CNFs are implemented in future explorations.

We note that these results demonstrate the feasibility of exact posterior sampling for Bayesian normalizing flows, leading to constraints that are more informative than the prior, and show that well-calibrated UQ and potential model comparison between architectures is achievable through this approach. They further illustrate how the quantified uncertainties can mitigate against biases that might result from un-marginalized variance in flow predictions. In addition, they demonstrate the data augmentation capabilities of flow-based PopSynth emulators and how UQ can enable reliable feature prediction in rarely synthesizable sub-populations without having to simulate a prohibitively large number of systems. The fully scalable implementation of this method using Bayesian continuous normalizing flows with robust convergence studies is left as a future exploration.

\subsection{Uncertainty quantification in density evaluation using Bayesian flows}

\begin{figure*}[htt]
\begin{center}
\includegraphics[width=0.96\textwidth]{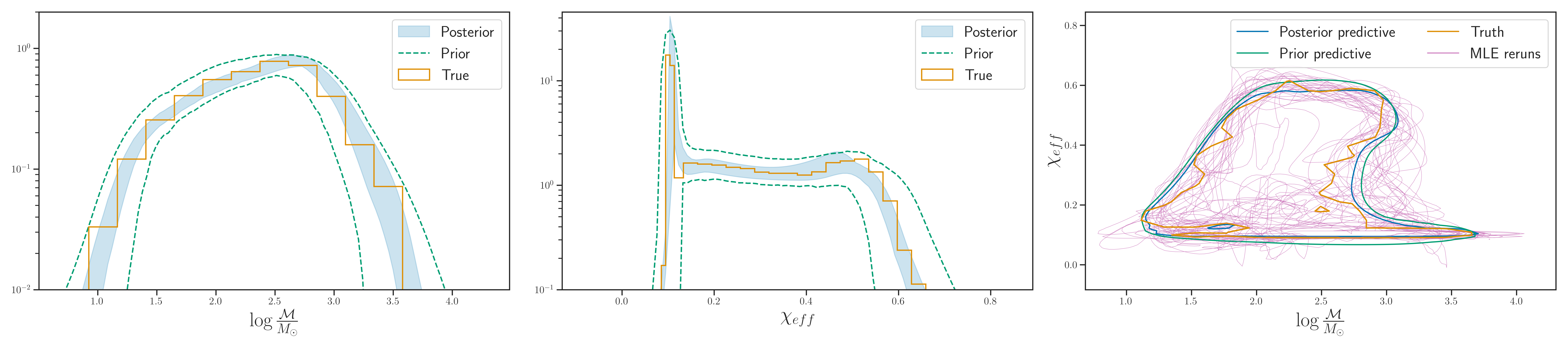}
\caption{\label{fig:uq}Predictions of the Bayesian NF for prior scale $\sigma_0=0.25$. Highest posterior density credible intervals on the marginal distribution~(left), and the predictive joint distribution~(right) for BBH chirp-mass and effective inspiral spin, are shown. }
\end{center}
\end{figure*}

To visualize UQ in density evaluation, we emulate the two-dimensional population of BBH chirp masses and effective inspiral spins, since higher-dimensional densities are difficult to represent and marginalize over grids. Bayesian inference of the flow parameters yields the posterior predictive distribution of the BBH population that comprises the test set, which is compared with the truth and the prior in Fig.~\ref{fig:uq}. It can be seen that the credible interval of the predicted population enclosed the true distribution, with the posterior being significantly more informative than the prior. Also, it can be seen that the posterior predictive distribution is much closer to the truth than individual MLE reruns on different realizations of the thinned dataset, while also outperforming the prior predictive distribution. This is indicative of the fact that for sparse training data, the posterior predictive emulator should be used for both SBI and feature prediction instead of MLEs. 

\subsection{Uncertainty quantification in sample generation using Bayesian flows}
\begin{figure*}[htt]
\begin{center}
\includegraphics[width=0.96\textwidth]{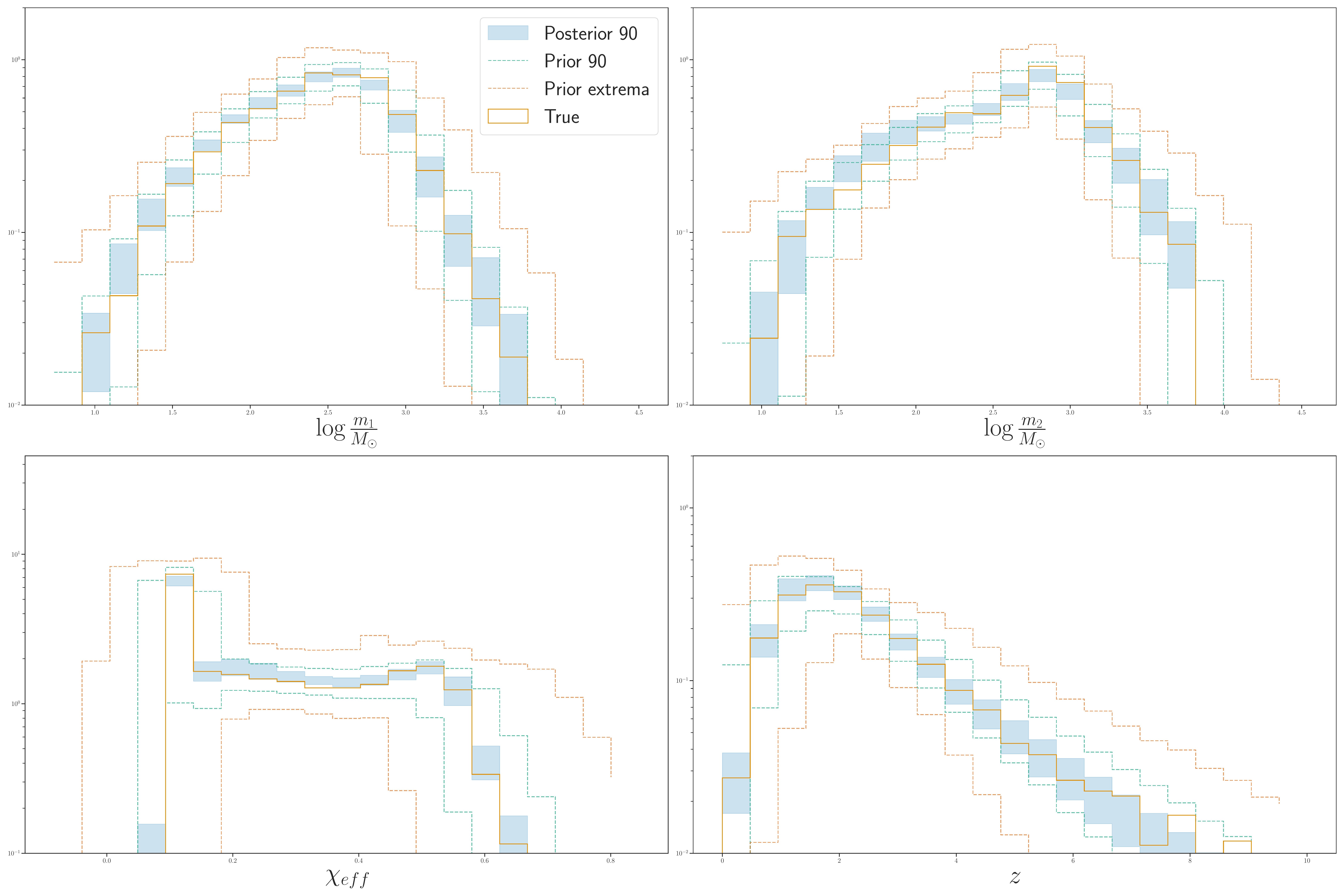}
\caption{\label{fig:uq4p} Uncertainty quantification for sample generation using Bayesian flows, for emulating the joint distribution of four BBH parameters}
\end{center}
\end{figure*}
For sample generation, it is straightforward to represent and marginalize higher-dimensional densities since the latter amounts to ignoring samples of the marginalized dimensions. Hence, for this demonstration, we emulate the four-dimensional distribution of BBH component masses, effective spins, and redshifts~(exactly similar to the emulators of \cite{colloms2025exploringevolutiongravitationalwaveemitters}). As before, Bayesian inference yields posterior samples of the flow parameters corresponding to each of which, we draw samples of BBH parameters and represent their densities as histograms. Figure~\ref{fig:uq4p} shows the Bayesian credible intervals on the density of histogram heights in each bin, compared with the truth and the prior. As in the case of density evaluation, it can be seen that the Posterior is encloses the true distribution of the test set while also being considerably more informative than the prior. In other words, the proposed method correctly quantifies uncertainty in sample generation by NF emulators.
\subsection{Data augmentation for feature prediction with UQ}

\begin{figure*}[htt]
\begin{center}
\includegraphics[width=0.46\textwidth]{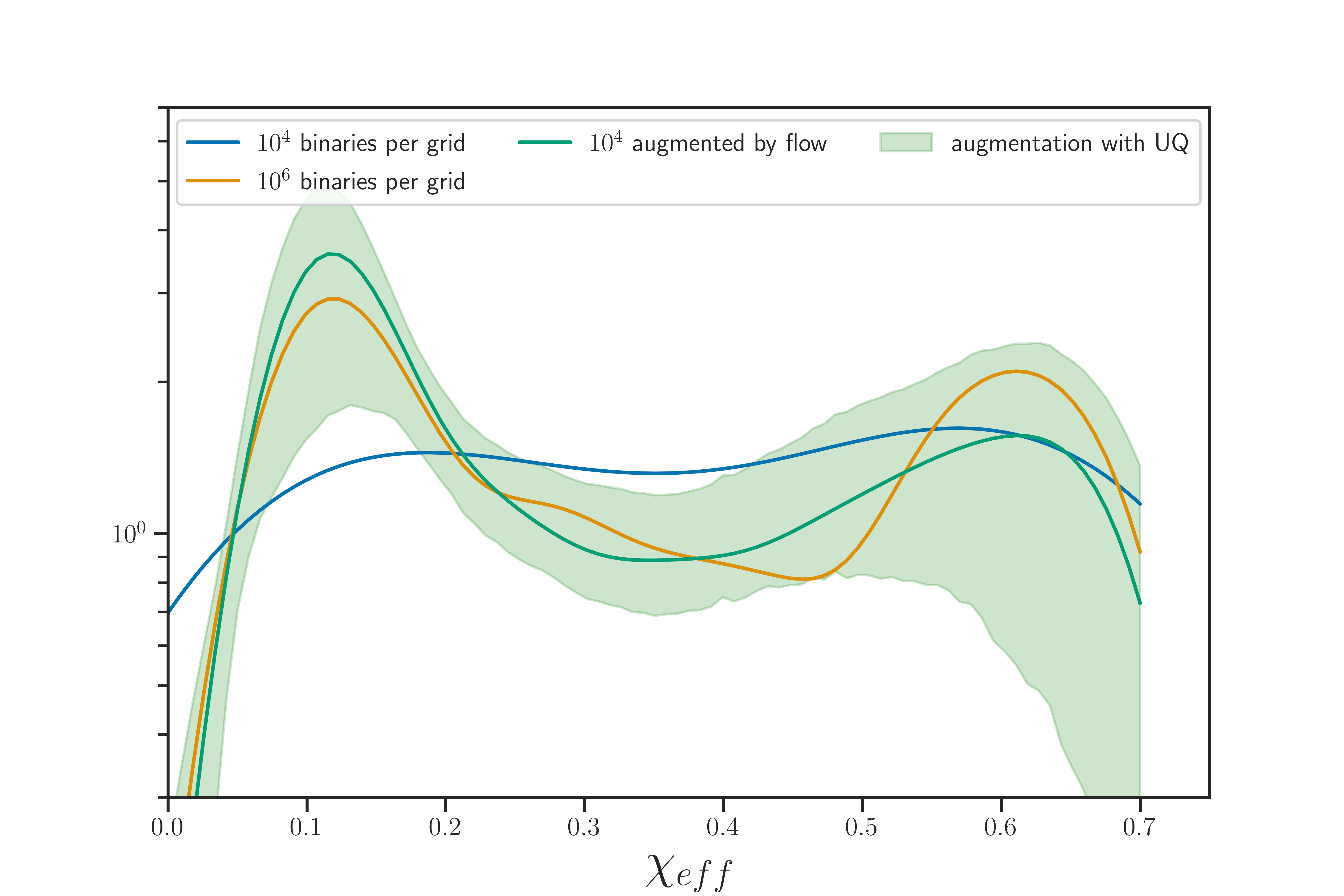}
\caption{\label{fig:bamp} Feature prediction with uncertainty quantification.}
\end{center}
\end{figure*}

As shown in Sec.~\ref{sec:methods:flows}, NF emulators can augment sparse PopSynth datasets to boost sample-statistics without having to simulate a prohibitive number of systems. However, training on small synthetic populations per grid point introduces uncertainties that need to be marginalized over to obtain accurate feature prediction~(even though data augmentation improves feature prediction, deviation from the truth is still observed in Fig.~\ref{fig:amp}, which can in principle be rectified by marginalizing over uncertainties). With Bayesian flows, we quantify these uncertainties in data augmentation by predicting a credible interval (an ensemble of augmented datasets) that encloses both the truth and the uncertain prediction from the MLE flow, as can be seen in figure~\ref{fig:bamp}.

\subsection{Model comparison}
To assess the efficiency of different flow architectures and avoid overfitting, we compute the BIC for our default architecture and compare it with that of a different~(more complex) one, both for the four-dimensional emulator, and list them as follows.
\vspace{-0.1cm}
\begin{center}
\setlength{\tabcolsep}{12pt} 
\renewcommand{\arraystretch}{1.5} 
\begin{tabular}{|c|c|c|c|}
\hline
Model & Architecture  &  $\log p_{max}(\vec{d}_{train}|\vec{\omega})$ & BIC \\
\hline
Default & 16 layers, 3 hidden layers, 150 neurons & 104  & 4519833 \\
\hline
Simpler & 17 layers, 3 hidden layers, 150 neurons &  -174  & 9377420 \\
\hline
\end{tabular}
\end{center}
\vspace{0.1cm}
Note that the high complexity of the MAF implies that the BICs are likely driven by the complexity term as compared to the maximum likelihood one. Nevertheless, it can be seen that the model with higher complexity also has a lower maximum log-likelihood achieved during posterior sampling. A dedicated and meaningful model comparison study for different architectures wit Bayesian CNFs, given various PopSynth datasets,  is left as a future exploration.

\subsection{Calibration of uncertainties}
\begin{figure*}[htt]
\begin{center}
\includegraphics[width=0.3\textwidth]{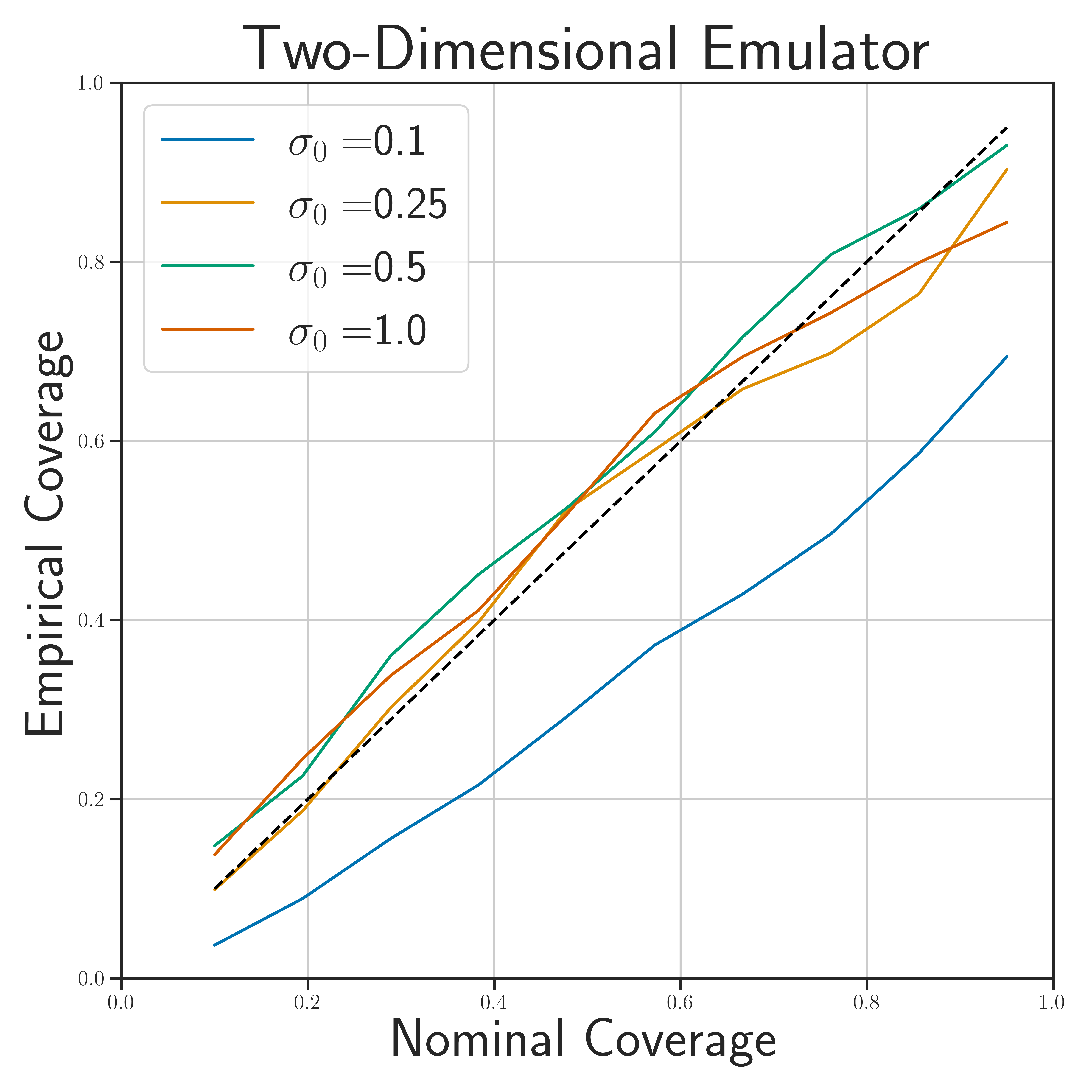}
\includegraphics[width=0.3\textwidth]{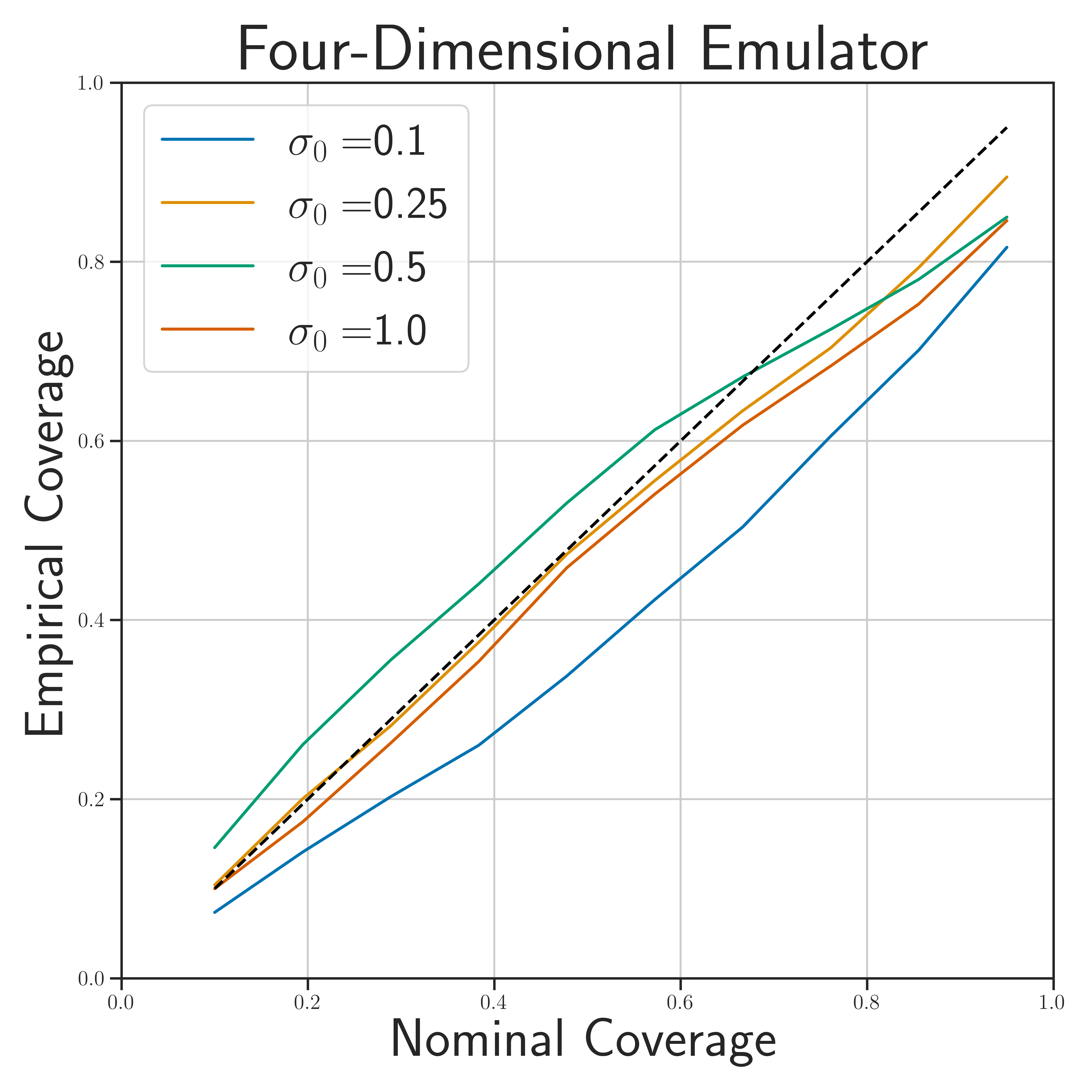}
\includegraphics[width=0.3\textwidth]{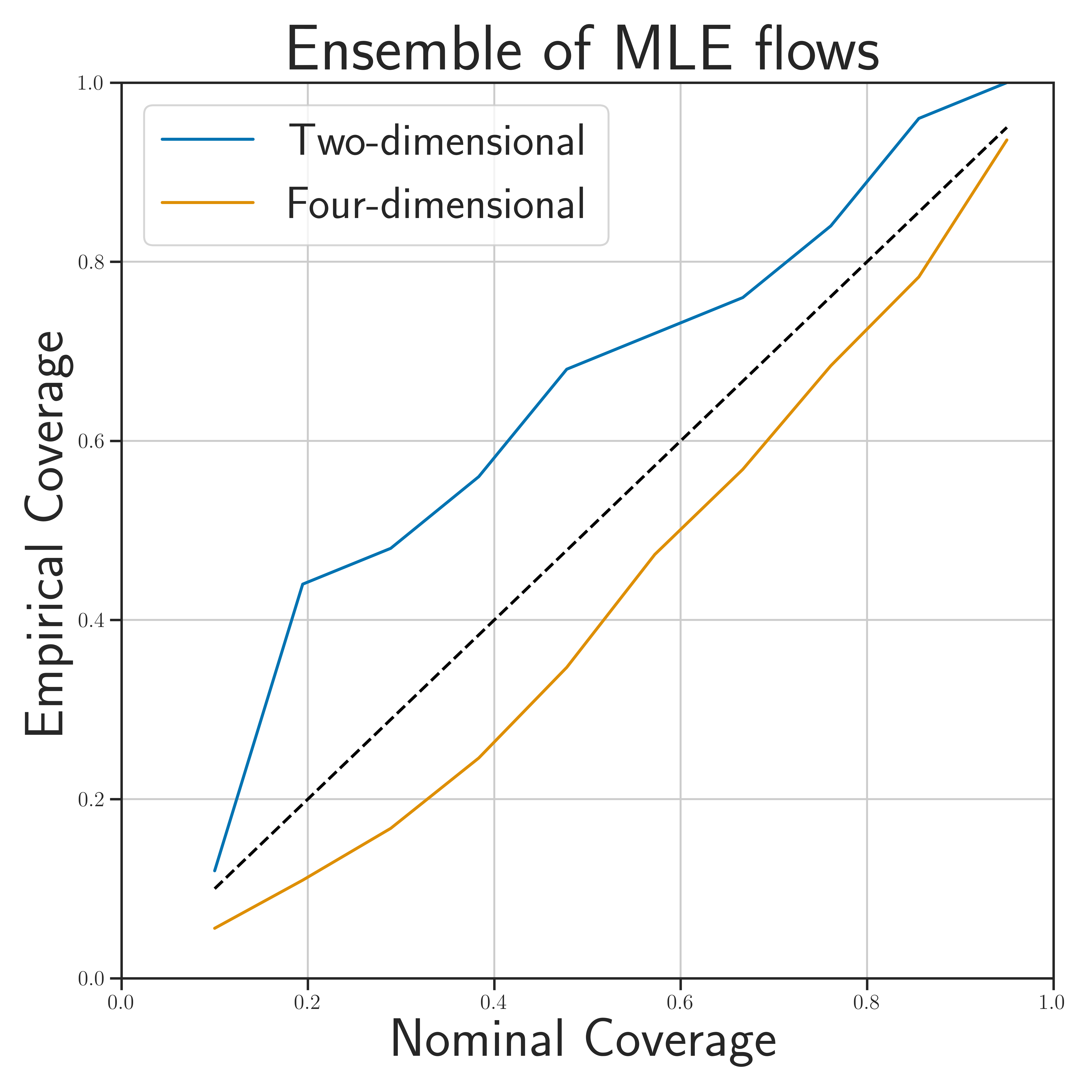}
\caption{\label{fig:calib} Calibration curves for Bayesian flows (left and center) withvarious choices of the prior scale $\sigma_0$ and an ensemble of flows trained using the MLE method~(right).}
\end{center}
\end{figure*}

To optimize the choice of $\sigma_{0}$ and assess the quantified uncertainties, we show in Fig.~\ref{fig:calib} calibration curves that compare the empirical coverage of flow predictions with the theoretical coverage for varying numbers of equal probability mass bins. For very small values of $\sigma_0$ the uncertainties are underestimated, which indicates that the inference is restricted by the prior. For increasing $\sigma_0$, the calibration curves improve, and slight over-estimation is observed for very high values of $\sigma_0$. The overestimation does not seem to increase too much with increasing $\sigma_0$, which is expected since broad priors are still consistent with accurate inference as long as there is support near the highest likelihood region. The observed overestimation can be interpreted as the sampler having difficulties converging for a heavily downsampled dataset, a highly complex model, and broad priors. For lower model complexity that guarantees scalable and convergent inference even for broad priors, higher values of $\sigma_0$ above a certain threshold are expected to have identical calibration curves. As mentioned before, very high values of $\sigma_0$ can also lead to all the walkers getting stuck at local maxima, causing biases, which the calibration curves would identify as underestimation of coverage.

Nevertheless, for the complex models used in this demonstration, good calibration is observed for $\sigma_0=0.25$, which is consistent with the credible intervals shown in figures~\ref{fig:uq}, \ref{fig:uq4p}. Furthermore, the Bayesian UQ outperforms the ensemble method used by \cite{Plunkett:2025mjr} which either over or under-estimates coverage for the test set. In other words, the estimated credible intervals indeed encompass the true population distribution the expected number of times, when averaged over all regions of the parameter space, justifying this choice of $\sigma_{0}$. Such well-calibrated Bayesian NFs will correctly quantify the uncertainties in PopSynth emulation from sparse datasets, with the posterior predictive emulators expected to enable unbiased SBI and feature prediction for \textsc{POSYDONv2} datasets, leading to reliable astrophysical insights from future GW catalogs. 

\section{Discussion and Future Prospects}
\label{sec:discussion}
In this work, we developed a self-consistent method for quantifying epistemic and aleatoric uncertainties in the predictions of NF models in the context of PopSynth emulators and potentially for selecting between different flow architectures to reduce overfitting. By relying on Bayesian neural networks to construct flow transformations and the natural log likelihood function associated with probabilitic models such as NFs, we have shown that well-calibrated UQ and plausible model comparison are achievable. Our uncertainty estimates can successfully mitigate the biases resulting from variance in flow predictions among multiple re-runs with different parameter initializations or alternative realizations of a sparse and noisy training sets, while also being better calibrated than the uncertainty estimates obtained from the ensemble. We have demonstrated this in the context of PopSynth simulations of merging BBHs, whose emulation is necessary for insightful astrophysical inference from observed catalogs. We have further shown that NF emulators equipped with UQ can accurately augment sparse PopSynth datasets for feature prediction in rarely synthesizable sub-populations without having to simulate a prohibitively large number of systems.

The developed methodology will be necessary for marginalizing over model uncertainties, which are expected to be non-negligible for sparse PopSynth datasets, which might be unavoidable for high fidelity~(and hence costly) simulators such as \textsc{POSYDONv2} or rarely synthesized sub-populations such as extreme mass-ratio binaries with compact objects in the purported lower-mass gap, and high metallicity binary stars that lead to very few merging BBH systems. In an ongoing investigation, we are exploring the performance of the developed emulators on a diverse collection of \texttt{POSYDONv2}, runs, which will facilitate robust simulation-based probes of binary stellar evolution from observed CBC catalogs, both employing direct SBI or through the guided development and inference of targeted phenomenological population models. Novel astrophysical insights free of biases arising from emulator uncertainties will be plausible.

We have discussed several planned improvements in terms of scalability that are necessary for fully realizing the potential of Bayesian NFs as a robust and tractable tool for UQ in simulation-based inference from observed astrophysical populations. These include the use of CNFs to reduce complexity~(appendix \ref{sec:appendix:cnf}), approximate Bayesian modeling using stochastic variational inference and importance sampling~(appendix \ref{sec:appendix:svi}), and alternative approaches such as NFs with Monte Carlo Dropouts~(appendix~\ref{sec:appendix:mcdp}). We have further alluded to more general applications of Bayesian NFs in  UQ for deep predictive and generative modeling in the context of likelihood-free parameter estimation~(appendix~\ref{sec:appendix:NLE}) with potential applications to other simulation-based approaches in GW astronomy~\citep{Dax_2021, Dax:2024mcn}, implementations of which are also left as future explorations. As we enter the era of large datasets, high fidelity simulators, and powerful deep learning models capable of combining the two into new scientific discoveries, fully realized Bayesian CNFs will play a crucial role in mitigating biases arising from un-marginalized model uncertainties and over/under fitting arising from sub-optimal architecture choices, there by facilitating reliable and accurate astrophysical inference and feature prediction.

\section{acknowledgements}
We are thankful to Vicky Kalogera for her insights and support throughout this project. We are grateful to Mathew Mould for a very detailed, informal review of the manuscript. We thank Aggelos Katsaggelos, Santiago Lopez Tapia, Ryan Mgee, Deep Chatterjee, Tri Nguyen, Kyle Rocha, Ugur Demir, and Philipp Srivastava for valuable discussions and inputs. This work was supported by the National Science Foundation~(NSF) award PHY-2207945. We gratefully acknowledge the support of the NSF-Simons AI-Institute for the Sky (SkAI) via grants NSF AST-2421845 and Simons Foundation MPS-AI-00010513. This research was also supported in part through the computational resources and staff contributions provided for the Quest high-performance computing facility at Northwestern University, which is jointly supported by the Office of the Provost, the Office for Research, and Northwestern University Information Technology. This research used the DeltaAI advanced computing and data resource, which is supported by the National Science Foundation (award OAC 2320345) and the State of Illinois. DeltaAI is a joint effort of the University of Illinois Urbana-Champaign and its National Center for Supercomputing Applications. We are grateful for the computational resources provided by the LIGO laboratory and supported by National Science Foundation Grants PHY-0757058 and PHY-0823459.
\software{Pytorch, Pyro, JAX, Numpyro
          }


\appendix
\section{Improving scalability}
\subsection{MAF vs CNF: complexity and expressivity}
\label{sec:appendix:cnf}
\begin{figure*}[htt]
\begin{center}
\includegraphics[width=0.46\textwidth]{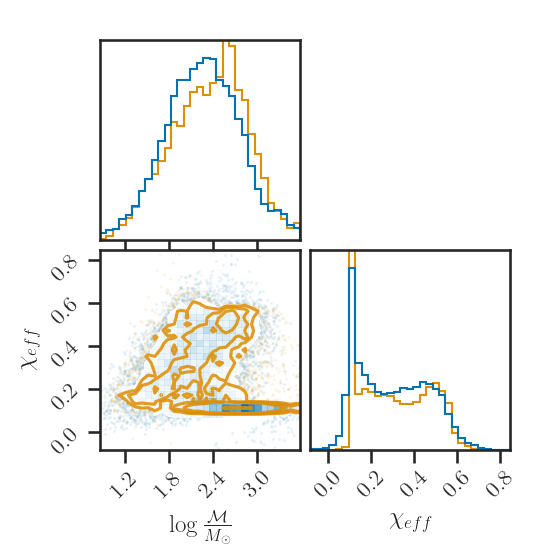}
\includegraphics[width=0.46\textwidth]{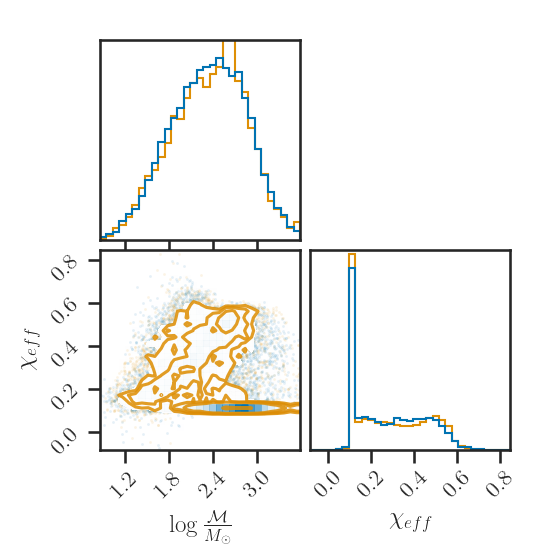}
\caption{\label{fig:cnf} Performance of a simple MAF(left) vs CNF(right), showing the emulated population~(blue) on the test set compared with the true distribution~(orange).}
\end{center}
\end{figure*}
As mentioned before, CNF algorithms can estimate densities with DNNs of much smaller complexity than MAFs. In Fig.~\ref{fig:cnf}, we show the performance of a much simpler MAF architecture~(2 autoregressive layers, each with 4 hidden layers spanned by 150 units) than the default one, along with a CNF~(trained using the FFJORD algorithm,~\cite{gal2016dropoutbayesianapproximationrepresenting}) which has a multi-layered perceptron composed of just three 64-unit layers for its neural network. The complexity~(number of free parameters) of the CNF is hence orders of magnitude smaller than even this simpler MAF. It can be seen how the simpler MAF struggles to estimate the conditional densities, whereas the CNF performs much better. This implies that a Bayesian CNF emulator for PopSynth will involve sampling a posterior of at most $O(10^4)$ parameters, which is well within the scope of NUTS. Hence, emulating higher-dimensional BBH populations, the use of broad and flexible priors will be feasible and consistent with convergent Bayesian inference, once the Bayesian CNF is implemented into production. Such a development is left as a future exploration.
\subsection{Scalable approximations: stochastic variational inference}
\label{sec:appendix:svi}
Even with the promise of scalable and convergent sampling with Bayesian CNFs, approximate Bayesian UQ methods can be useful for scenarios wherein CNFs with simple DNNs are not an option or require complex embedding networks for $\vec{\lambda}$, which is usually the case in simulation-based inference problems that do not involve emulating PopSynth, but rather approximating the intractable likelihood of observable parameters given high-dimensional data. If the dimensionality of model parameters cannot be reduced from $O(10^6)$ or higher, scalable alternatives to posterior sampling should be considered.

In stochastic variational inference~\citep[SVI,][]{hoffman2013stochasticvariationalinference}, the posterior distribution of model parameters $p(\vec{\omega}|\vec{d}_{train})$ is approximated using some fiducial distribution $Q(\vec{\omega}|\vec{\Omega}_q)$ such as a mixture of truncated Gaussians, characterized by its own set of parameters $\vec{\Omega}_q$. Optimal values of these parameters can be learned by minimizing the evidence-based lower-bound between $p(\vec{\omega}|\vec{d}_{train})$ and $\vec{\Omega}_q$, which can be computed using:
\begin{equation}
    \mathrm{ELBO}(\vec{\Omega}_q) = \frac{1}{N_{q}}\sum_{\vec{\omega}_i\sim Q(\vec{\omega}|\vec{\Omega}_q)}\left\{\log p(\vec{\omega}_i|\vec{d}_{train}) - \log Q(\vec{\omega}_i|\vec{\Omega}_q) \right\}
\end{equation}
where $N_q$ is the number of $\vec{\omega}$ samples drawn to estimate the ELBO. Given optimal parameters $\vec{\Omega}_{q,0} $ obtained using gradient descent algorithms such as \textsc{Adam}~\citep{kingma2017adammethodstochasticoptimization}, samples of $\vec{\omega}\sim Q(\vec{\omega}|\vec{\Omega}_q)$ can be obtained, which can be re-weighted to the true posterior using importance sampling. The weights~$(w_i)$ can be used to compute the effective sampling size~(ESS), which can be used as a reliability test. They can further be used to compute the Bayesian evidence~$(Z)$ of the flow model, which can in turn be used for model comparison among architectures:

\begin{equation}
    w_i = \frac{p_{\omega}(\vec{\omega}_i)p(\vec{d}_{train}|\vec{\omega})}{Q(\vec{\omega}_i|\vec{\Omega}_{q,0})}
\end{equation}
\begin{equation}
    \mathrm{ESS} = \frac{\sum_i w_i^2}{\left(\sum_j w_j\right)^2}
\end{equation}
\begin{equation}
    Z = \frac{1}{N}\sum_i w_i
\end{equation}

This approach towards UQ retains the Bayesian interpretation of credible intervals and can usually be considered reliable provided the ESS after importance-sampling is close to the original number of samples drawn from $Q$. In certain cases however, sub-optimal prior choices can lead to poor uncertainty estimation even for high ESS. In such a scenario, variants of the importance sampling methods such as the ones used by~\citep{mould2025rapidinferencecomparisongravitationalwave} could be investigated. A detailed study of SVI vs exact Bayesian inference for UQ in NF predictions is left as a future exploration. Nevertheless, this approximate Bayesian approach can be expected to be fully scalable for complex flow architectures.

\subsection{Scalable alternatives: Monte Carlo dropout flows}
\label{sec:appendix:mcdp}
An alternative to Bayesian UQ by means of direct posterior sampling or its more scalable SVI-based counterpart, is the implementation of Monte Carlo dropout while training and evaluating the NN underlying an NF model. In this approach, a certain fraction of hidden units in each layer of an NFs network, chosen at random, are deactivated during MLE training. In other words, the trained flow comprises an ensemble of smaller architectures, each capable of approximating the target distribution with reasonable accuracy~\citep{berry2023normalizingflowensemblesrich}. During prediction, instead of turning off dropout, it is allowed to function as it were during training, leading to an ensemble of predictions each corresponding to a single member of the mentioned collection of smaller architectures. Constructing confidence intervals from this ensemble of predictions can be used to quantify epistemic uncertainties, perhaps even in the Bayesian sense, as shown in previous studies for deterministic NN models~\citep{gal2016dropoutbayesianapproximationrepresenting}. However, an equivalent derivation for NFs, as well as the scope of MC dropout NFs in quantifying aleatoric uncertainties such as the ones PopSynth emulators can be susceptible to, is left as a future exploration.

\section{Applications beyond PopSynth}

\subsection{Likelihood free inference marginalized over model uncertainties}
\label{sec:appendix:NLE}
In addition to population-level probes of observed catalogs, NFs have been used in the literature to approximate intractable likelihoods and thereby enable fast and scalable Bayesian modeling given simulated datasets~\citep{papamakarios2019sequentialneurallikelihoodfast}, with CBC parameter estimation applications~\citep{Dax_2021,Dax:2024mcn}. By simulating the observable data $\vec{x}$ corresponding to parameters $\vec{\theta}$ and accurate assumptions on the distribution of detector noise~$(\vec{n})$, NFs can be trained to learn the posterior distribution $p(\vec{\theta}|\vec{x})$, which can be used for rapid parameter estimation from newly observed data. Amortized inference can be achieved by concatenating the data with the simulated noise, should the noise be reliably measurable during observational inference. Embedding networks are often used to compress the data for feature extraction and data reduction. The estimator would take the following form:
\begin{equation}
    \hat{p}(\vec{\theta}|\vec{x},\vec{n},\vec{\omega}) = p_{u}(\vec{u}=f_{NN}(\vec{\theta},g_{NN}(\vec{x}, \vec{n}, \vec{\omega}_e),\vec{\omega}_f))\left|\frac{\partial\vec{u}}{
    \partial \vec{\theta}}\right|.
\end{equation}
where $g_{NN}$ is the embedding network and the model parameters $\vec{\omega} = (\vec{\omega}_{f},\vec{\omega}_e)$  now comprise both the parameters of the flow~$\vec{\omega}_f$ and the embedding net~$\vec{\omega}_{e}$. The estimator function can be used to construct the likelihood of the model parameters $p(\vec{d}_{train}| \vec{\omega})$ similar to Eq.~\eqref{eq:loglike} given training data: $\vec{d}_{train} = \{(\vec{\theta}_i, \vec{x}_i, \vec{n}_i)\sim p(\vec{x}|\vec{\theta}, \vec{n})p(\vec{n})p(\vec{\theta}) \}$, where $p(\vec{\theta})$ is a prior used for generating the training points, $p(\vec{n})$ is a model for the noise distribution and samples from the intractable distribution $p(\vec{x}|\vec{\theta}, \vec{n})$ is representative of running the simulator for every grid point and adding noise to the generated data. Once the training data is generated and the likelihood constructed, Bayesian UQ and model comparison can be implemented using the framework presented in this paper. Bayesian CNFs and SVI will likely be required for problems with high-dimensional parameters, data, and complex embedding networks, such as real-time GW parameter estimation~\citep{Dax_2021, Dax:2024mcn}.



\bibliography{sample7}{}
\bibliographystyle{aasjournalv7}



\end{document}